\documentstyle[epsfig]{l-aa}
\include{psfig}
\begin{document}
\thesaurus{ (04.19.1) (08.22.1) (11.13.1)}
\title{A slope variation in the Period-Luminosity relation for short period
SMC Cepheids \thanks{Based on observations made at ESO, La Silla, Chile.}
}
\author{
F.~Bauer\inst{1,2}, C.~Afonso\inst{2}, J.N.~Albert\inst{3}, C.~Alard\inst{4},
J.~Andersen\inst{5}, R.~Ansari\inst{3}, {\'E}.~Aubourg\inst{2},
P.~Bareyre\inst{1,2}, J.P.~Beaulieu\inst{4,12},
A.~Bouquet\inst{1},
S.~Char\inst{6}$^{\dagger}$, X.~Charlot\inst{2}, F.~Couchot\inst{3}, C.~Coutures\inst{2},
F.~Derue\inst{3}, R.~Ferlet\inst{4}, C.~Gaucherel\inst{2},
J.F.~Glicenstein\inst{2}, B.~Goldman\inst{2,7,8},
A.~Gould\inst{9}\thanks{Alfred P. Sloan Foundation Fellow},
D.~Graff\inst{2,10}, M.~Gros\inst{2}, J.~Haissinski\inst{3},
J.C.~Hamilton\inst{1}, D.~Hardin\inst{2}, J.~de Kat\inst{2}, A.~Kim\inst{1},
T.~Lasserre\inst{2}, {\'E}.~Lesquoy\inst{2}, C.~Loup\inst{4},
C.~Magneville \inst{2}, B.~Mansoux\inst{3}, J.B.~Marquette\inst{4},
{\'E}.~Maurice\inst{11}, A.~Milsztajn \inst{2}, M.~Moniez\inst{3},
N.~Palanque-Delabrouille\inst{2}, O.~Perdereau\inst{3}, L.~Pr{\'e}vot\inst{11},
C.~Renault\inst{2}, N.~Regnault\inst{3}, J.~Rich\inst{2},
M.~Spiro\inst{2}, A.~Vidal--Madjar\inst{4},
L.~Vigroux\inst{2}, S.~Zylberajch\inst{2} -- The EROS collaboration
}
\institute{
     Coll{\`e}ge de France, PCC, IN2P3 CNRS, 11 place Marcelin Berthelot, 75231 Paris Cedex, France
\and CEA, DSM, DAPNIA, Centre d'{\'E}tudes de Saclay, 91191 Gif-sur-Yvette, Cedex, France
\and Laboratoire de l'Acc{\'e}l{\'e}rateur Lin{\'e}aire, IN2P3 CNRS et Universit{\'e} Paris-Sud, BP~34 91898 Orsay Cedex, France
\and Institut d'Astrophysique de Paris, INSU CNRS, 98~bis Boulevard Arago, 75014 Paris, France
\and Astronomical Observatory, Copenhagen University, Juliane Maries Vej 30, 2100 Copenhagen, Denmark
\and Universidad de la Serena, Facultad de Ciencias, Departamento de Fisica, Casilla 554, La Serena, Chile
\and Dept. Astronom{\'\i}a, Universidad de Chile, Casilla 36-D, Santiago, Chile
\and European Southern Observatory, Casilla 19001, Santiago 19, Chile
\and Department of Astronomy, Ohio State University, Columbus, OH 43210, U.S.A.
\and Physics Department, Ohio State University, Columbus, OH 43210, U.S.A.
\and Observatoire de Marseille, 2 place Le Verrier, 13248 Marseille Cedex 04, France
\and Kapteyn Laboratorium, Postbus 800, 9700 AD Groningen, Netherlands
}
\offprints{Florian.Bauer@cea.fr}
\date{Received xx xx, 1998; accepted xx xx, 1999}
\maketitle
\markboth{Bauer et al. 1999: A slope variation in the PL 
relation of short period SMC Cepheids
}{}
\begin{abstract}
We present the Period--Luminosity relations 
from 290 Cepheids towards the LMC and 590 Cepheids towards the SMC.
The two data sets were obtained
using the two wide field CCD cameras of the EROS~2 microlensing survey.
We observe a significant slope change of the period--luminosity relation for
the SMC fundamental mode Cepheids with periods shorter than 2 days.
Many possible experimental biases have been investigated, but none
can account for this effect.
We also observe different spatial distributions for SMC Cepheids with different
ages i.e periods.
Different possible explanations of the slope change are discussed.
\keywords{Stars: Surveys -- Stars: Cepheids -- Galaxies: Magellanic Clouds} 
\end{abstract}
\section{Introduction}
The Cepheid period--luminosity (PL) relation has been one of the cornerstones
of distance determination since the beginning of this century.
Most of the recent determination of  the Hubble constant $H_0$
are based on HST observations of Cepheids as far away as the Virgo and
Fornax clusters (\cite{ferrarese}; \cite{sandage}; \cite{tanvir}),
an assumed universal PL relation calibrated in the LMC,
and the so-called \cite{madore} method.
At the same time, modern evolutionary and pulsation calculations are
under way to obtain a theoretical calibration
of the Cepheid PL relation (\cite{baraffe}; \cite{wood};
\cite{alibert}; \cite{bono99}; 
%\cite{saio}; 
\cite{chiosi}; \cite{yecko}; \cite{buchler}).

The number of available Cepheids towards the Magellanic Clouds has been
dramatically increased by
the different microlensing surveys (e.g. \cite{beaulieu96}; \cite{welch};
\cite{beaulieu97a}).
With the commissionning of the new EROS survey equipment
(EROS~2, Bauer et al. 1997), a new systematic search
for Cepheid variables in the LMC and SMC was undertaken in October 1996.
In this article, we present the first result from this search, a
so far unreported
slope change of the PL relation for fundamental mode
SMC Cepheids at short periods.

We describe the observational setup and data reduction in Sect.~2.
In Sect.~3 we present the PL relations of fundamental mode (hereafter F)
and first overtone (hereafter 1-OT) Cepheids towards
both Magellanic Clouds.
The PL relation of F Cepheids towards the SMC displays a visible non--linearity
for periods shorter than about 2 days, and we establish
the statistical significance of this effect.
We then discuss different possible observational biases. 
In Sect.~4 we study the spatial distribution
and in Sect.~5 we present the depth dispersion of our SMC Cepheids.
In Sect.~6 we develop possible scenarios that could explain the effect,
before summarizing in Sect.~7.
\section{Observations and data reduction}
We give here only a brief overview as more details will be available
in the EROS~2 Cepheid catalog (\cite{bauer3}).
Observations were obtained using the new EROS~2 experimental setup,
which consists of a 1~m Ritchey--Chretien
telescope and two 4k~$\times$ 8k CCD mosaic cameras,
allowing simultaneous
imaging in two focal planes with different colour passbands
(420--650 nm, so called EROS-2 $V_{ \rm EROS}$ passband ; and
650--900 nm, so called EROS-2 $R_{ \rm EROS}$  passband)\footnote{We stress that
these filters are different from those of the EROS~1 programme (see e.g.
\cite{sasselov_metal}, \cite{grison}).}.
Each CCD mosaic is made up of eight 2k~$\times$ 2k three edge buttable
thick CCDs developped by Loral/U.Arizona.
The pixel size is 0.6~arcsec (15~$\mu$m). The available global
field of view is 0.7$^{\circ}$ (right ascension)~$\times$ 1.4$^{\circ}$ (declination).
The whole system was mounted at the La Silla Observatory in Chile
in June 1996; details can be found in \cite{bauer1} and \cite{bauer2}.

Between October 1996 and February 1997,
a dedicated Cepheid campaign was undertaken.
Two fields per Magellanic Cloud (see Table \ref{table_champ}) were monitored
about once per night with an exposure time of 20 seconds.
Whenever possible, the four fields were imaged almost simultaneously
at the same airmass.
A total of $\sim$110-160 images was obtained for each field.
\begin{table}
\begin{tabular}{l|c|c|c|c}
\hline
field            &   LMC1     &  LMC2       &  SMC1      &  SMC2     \\
\hline
$\alpha$         &  5 23 34   &  5 15 36    &  0 51 54   &  0 41 54  \\
$\delta$         &-69 44 22   &-69 44 22    &-73 36 32   &-73 42 32  \\ 
\hline
\end{tabular}
   \caption{Coordinates (J2000) of the field centres. }\label{table_champ}
\end{table}
After photometric reduction using the standard EROS~2
photometry package (Peida++, \cite{ansari1}), we obtained
1,134,000 and 504,000 light curves of stars towards the LMC
and SMC, respectively. 
The relative inter--CCD
calibration of these light curves has an accuracy of 0.04~mag.\\
In the following presentation we used only $2\times6$ out of $2\times8$ CCDs per field.
One red CCD ceased to function in June 1996;
Cepheid detection would still be possible on the blue counterpart CCD, but no color
would be available. Another red CCD had a variable offset in time, so that a dedicated
treatment would be necessary in order to get proper magnitudes for the detected
Cepheids. This will be done in the future. As our main goal was to get
a clean set of Cepheid data we concentrated our efforts on the remaining 
$2\times6$ CCDs.
\begin{figure}[h]
\hbox{\epsfig{file=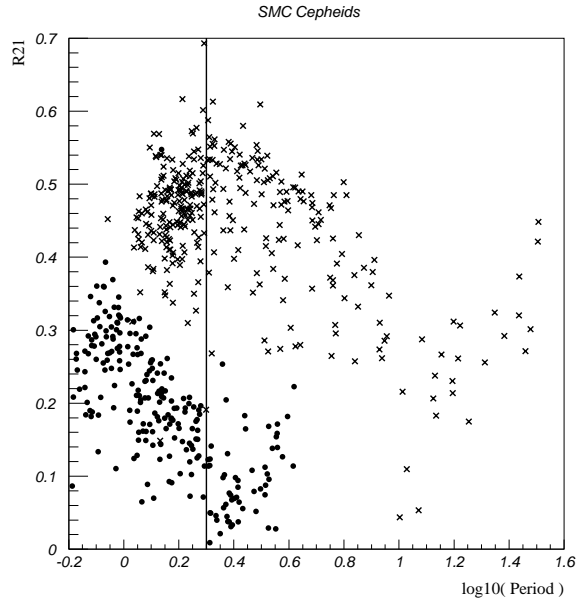,width=8.5cm} }  
\caption{The $R_{21}-\log P$ diagram for SMC F~Cepheids (crosses) and 1-OT Cepheids 
(black dots). The vertical line indicates the slope change of the PL-relation 
at 2 days for F~Cepheids.}\label{r21}
\end{figure}

A systematic search for periodic variable stars
was then performed (for $P > 0.5\,$ day),
using an algorithm  proposed by \cite{scargle}.
We fitted a 5$^{th}$-order Fourier series to the light curves;
in the following, the reported magnitudes are intensity--weighted
mean magnitudes\footnote{The magnitude that corresponds to the
average luminosity, as opposed to the average magnitude.} from this Fourier analysis.
Cepheid candidates were extracted using loose selection criteria in
both the Colour--Magnitude (CM) and PL diagrams, excluding population II Cepheids.
These cuts selected about 0.1~\% of the stars,
that were all inspected visually to reject non--Cepheid variables
(about one third of the scanned sample).
Within this final Cepheid sample, Fourier coefficients allow to distinguish
F Cepheids from 1-OT Cepheids, as was done in \cite{beaulieu95}
following the suggestion of \cite{antonello86}: 1-OT Cepheids
have a lower content of second and third harmonics than F~Cepheids.
Fig. \ref{r21} illustrates how this separation can be done, using
the amount of second harmonic, R$_{21}$, as a function of the pulsation period.  
A second visual inspection of Cepheid light curves 
was then performed for the few stars that lie in-between
the F and 1-OT Cepheid samples.
\begin{figure*}
\hbox{\epsfig{file=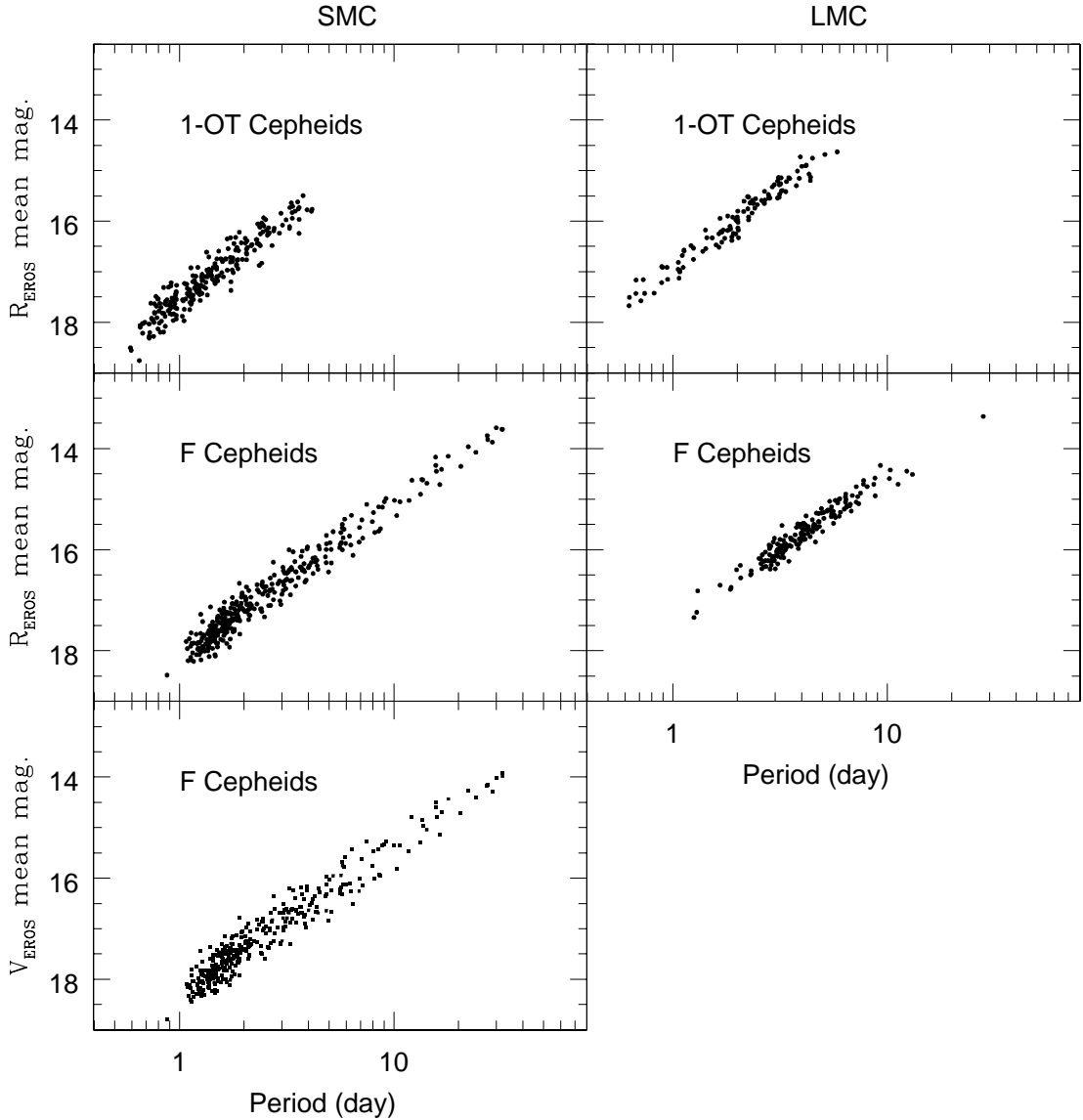,width=16.0cm}}
\caption{PL relations for first overtone
Cepheids (top) and fundamental mode Cepheids (center)
towards the SMC (left) and the LMC (right)
seen in R$_{ \rm EROS}$. The figure at bottom left shows the PL
relation for SMC F Cepheids seen in V$_{ \rm EROS}$. The five
panels are drawn to the same
scale to facilitate comparison.}\label{pl}
\end{figure*}
As a result we obtained a new EROS~2 Cepheid catalog, 
that will be published in a forthcoming paper
(\cite{bauer3}), comprised of 590 Cepheids towards the SMC
(351 F and 239 1-OT Cepheids) and 290 Cepheids towards the LMC
(177 F and 113 1-OT Cepheids).
\begin{table*}[tbh]  
\caption{Parameters of the PL relation fits described in the text. 
N is the number of stars in the sample,
$\sigma_{ \rm res}$ the dispersion of the fit residuals,
$\beta$ the slope of the PL relation and 
$\alpha$ its offset at $P = 1$~day;
in method 2 $\alpha$ is obtained as 
$\alpha_{\rm break} + \beta_{\rm i} \; \log(P_{\rm break})$.
We caution the reader that the offsets $\alpha$ are given 
in the EROS~2 filter system, which is not standard.}
\begin{tabular}{c|c|l|c|c|c|c|c|c|c}
\hline
                &       &               &       & \multicolumn{3}{c|}{V$_{ \rm EROS}$} & \multicolumn{3}{c}{R$_{ \rm EROS}$} \\
\cline{5-10}
Population      & Cloud & ~~~Method   & N     & $\alpha$      & $\beta$ & $\sigma_{ \rm res}$ & $\alpha$     & $\beta$       & $\sigma_{ \rm res}$\\
\hline
                &               & 1: ~~~all        & 177   & 17.63 $\pm$ 0.04 & 2.77 $\pm$ 0.07       & 0.18  & 17.44 $\pm$ 0.04      & 2.89 $\pm$ 0.06 & 0.15 \\
                & LMC           & 2: P $>$ 2 d     & 170   & 17.60 $\pm$ 0.05 & 2.72 $\pm$ 0.08       & 0.17  & 17.43 $\pm$ 0.04      & 2.89 $\pm$ 0.06 & 0.14  \\
F Cepheids      &               & 2: P $<$ 2 d     & 7     & 17.86 $\pm$ 0.16 & 3.59 $\pm$ 0.54       & 0.18  & 17.46 $\pm$ 0.13      & 2.99 $\pm$ 0.45 & 0.17  \\
\cline{2-10}
                &               & 1: ~~~all        & 351   & 18.29 $\pm$ 0.02 & 2.91 $\pm$ 0.04       & 0.24  & 18.12 $\pm$ 0.02      & 3.04 $\pm$ 0.03 & 0.20  \\
                & SMC           & 2: P $>$ 2 d     & 164   & 18.20 $\pm$ 0.04 & 2.80 $\pm$ 0.05       & 0.25  & 18.05 $\pm$ 0.03      & 2.95 $\pm$ 0.05 & 0.20  \\
                &               & 2: P $<$ 2 d     & 187   & 18.40 $\pm$ 0.04 & 3.48 $\pm$ 0.19       & 0.22  & 18.21 $\pm$ 0.04      & 3.49 $\pm$ 0.16 & 0.20  \\
\hline
                &               & 1: ~~~all        & 113   & 17.11 $\pm$ 0.03 & 3.12 $\pm$ 0.08       & 0.18  & 16.93 $\pm$ 0.03      & 3.18 $\pm$ 0.07 & 0.16  \\
                & LMC           & 2: P $>$ 1.4 d   & 87    & 17.13 $\pm$ 0.05 & 3.15 $\pm$ 0.12       & 0.19  & 16.96 $\pm$ 0.04      & 3.23 $\pm$ 0.11 & 0.16  \\
1-OT Cepheids   &               & 2: P $<$ 1.4 d   & 26    & 17.11 $\pm$ 0.03 & 3.03 $\pm$ 0.23       & 0.16  & 16.93 $\pm$ 0.03      & 3.07 $\pm$ 0.20 & 0.16  \\
\cline{2-10}
                &               & 1: ~~~all        & 239   & 17.70 $\pm$ 0.02 & 3.07 $\pm$ 0.08       & 0.25  & 17.55 $\pm$ 0.02      & 3.21 $\pm$ 0.08 & 0.21  \\
                & SMC           & 2: P $>$ 1.4 d   & 103   & 17.64 $\pm$ 0.05 & 2.90 $\pm$ 0.15       & 0.24  & 17.51 $\pm$ 0.04      & 3.07 $\pm$ 0.13 & 0.21  \\
                &               & 2: P $<$ 1.4 d   & 136   & 17.70 $\pm$ 0.02 & 3.27 $\pm$ 0.18       & 0.25  & 17.55 $\pm$ 0.02      & 3.37 $\pm$ 0.16 & 0.22  \\
\hline
\end{tabular}
\end{table*}\label{table_fit}
\section{Data analysis}
\subsection{Phenomenology}
The PL relations for the F and 1-OT Cepheids
are shown in Fig.~\ref{pl}.
For LMC Cepheids, and SMC 1-OT Cepheids as well, the relations are
compatible with a linear behaviour.
In contrast, that of the SMC F~Cepheids displays
a change in the slope for periods smaller than 2~days, visible in
both colours.
The magnitude deviation for these short--period Cepheids,
with respect to an
extrapolation of the PL relation for longer period
Cepheids, reaches 0.2~mag at $P = 1\,$ day.  Hence,
a simple linear fit to the full SMC F~Cepheid sample
would result in a biased slope and thus an incorrect estimate
of the distance to the SMC.
\subsection{Statistical significance of the effect}
In order to quantify the significance of the non--linearity in Fig.~\ref{pl},
we fit the PL relations in two different ways: 
(1) using a linear regression for the full data; 
(2) using two straight lines that cross at a ``break--period'' $P_{ \rm break}$,
\begin{eqnarray}
m(P) =  \alpha_{\rm break} - \beta_{ \rm i} \; \log(P / P_{ \rm break}) &  \ \ \ \ \  
\end{eqnarray}
where $i$ is an index for $ P < P_{ \rm break}$ (resp. $ P > P_{ \rm break}$).
We find the break--period to be 2.0 days for F~Cepheids and tentatively
use 1.4 day for 1-OT~Cepheids from the known ratio of F and 1-OT
Cepheid periods (see e.g. \cite{alcock95}, \cite{welch}, \cite{beaulieu97b}). 
The results from the fits are not sensitive to the removal of Cepheids
that lie close to the limit we defined to distinguish between 
F and 1-OT Cepheids.

As the PL relation scatter is mainly due to the width of the 
instability strip, differential reddening effects 
and depth dispersion, one cannot use the measurement errors as errors in 
the fits.
Instead, the errors used for the fits
are, for each Cepheid, the dispersion $\sigma_{ \rm res}$ from the 
standard linear regression (1).  Our data is compatible
with a constant dispersion whatever the Cepheid period, therefore 
we used an identical weight of $1 / \sigma_{ \rm res}^2$ for all Cepheids in a
given sample. The parameters from the fits are given in Table~\ref{table_fit}. 
Two results have to be pointed out:
first, the slope of the SMC F~Cepheids long--period sample 
from the fit of method (2) is closer to that
of the LMC F~Cepheid sample; 
second, the SMC F~Cepheids with $P < 2\,$ days
follow a PL relation with a much steeper slope.\\ 
We estimate the actual significance of the effect 
using the $\chi^2$ gain between fits (1) and (2). 
For the SMC F~Cepheids we calculate
a significance of  2.9~$\sigma$ in the R$_{ \rm EROS}$ passband
and 3.1~$\sigma$ in the V$_{ \rm EROS}$ passband, 
corresponding to a false detection probability of 0.2\%.
Thus, the slope change is statistically significant for SMC F~Cepheids.

We have also searched for a similar slope change in the SMC 1-OT 
Cepheids, but find no significant effect 
(0.9 $\sigma$ in R$_{ \rm EROS}$ and 1.3~$\sigma$ in V$_{ \rm EROS}$).
There is a caveat~: due to the smaller statistics and lever arm in $\log(P)$, we
expect less sensitivity to any possible slope change in this sample.
\subsection{Study of systematic biases}
Several tests or checks were carried out to determine whether this 
slope change is due to a systematic bias or not:
\begin{enumerate}
\item We exclude that uncertainties on the period lead to a bias:
where possible, comparison with existing catalogs (20 Cepheids)
shows that the determined period is accurate 
to better than 10$^{-3}$ day for a 2 days period. \\
\begin{figure}[htb]
 \hbox{\epsfig{file=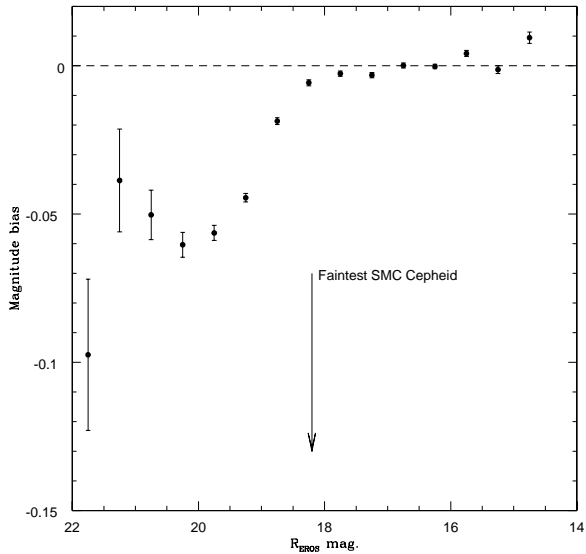,width=8.0cm} }    
\caption{Comparison of the stellar flux for images
from the current study (T$_{ \rm exp}$~=~20~s) 
with the stellar flux measured from images of the same field from the 
microlensing search (Palanque--Delabrouille et al. 1998; T$_{ \rm exp}$~=~300~s).
Stars with a mean magnitude brighter than 14.5 are omitted as they 
are saturated on the long exposure images.}\label{compa_sm}
\end{figure}
\item For stars with fluxes comparable to the sky background
our photometry is biased, such that the mean magnitudes 
could be underestimated.  
We have studied this 
by comparing the magnitudes of stars measured on 
images used for the present analysis, with those obtained
on a high quality image (good seeing and longer exposure time) 
of the same field.
Such images are available from our microlensing survey, and
have a 15 times longer exposure (300~s).
The result of this comparison is displayed
in Fig.~\ref{compa_sm}, which shows that stars at least as bright as 
the faintest SMC F~Cepheid have a bias smaller than 0.02~mag, much lower
than the observed deviation of 0.2~mag. \\
\item The non--linearity in the PL relation is visible for
F Cepheids but not for 1-OT Cepheids of the same magnitude. \\
\item 
We changed various parameters of our photometry package (PSF shapes, 
window size of the PSF adjustment etc.) and were unable to reproduce the observed
non-linearity. Thus the change in the slope is independent of our
photometry package.
\end{enumerate}
\begin{figure}[h]
\hbox{\epsfig{file=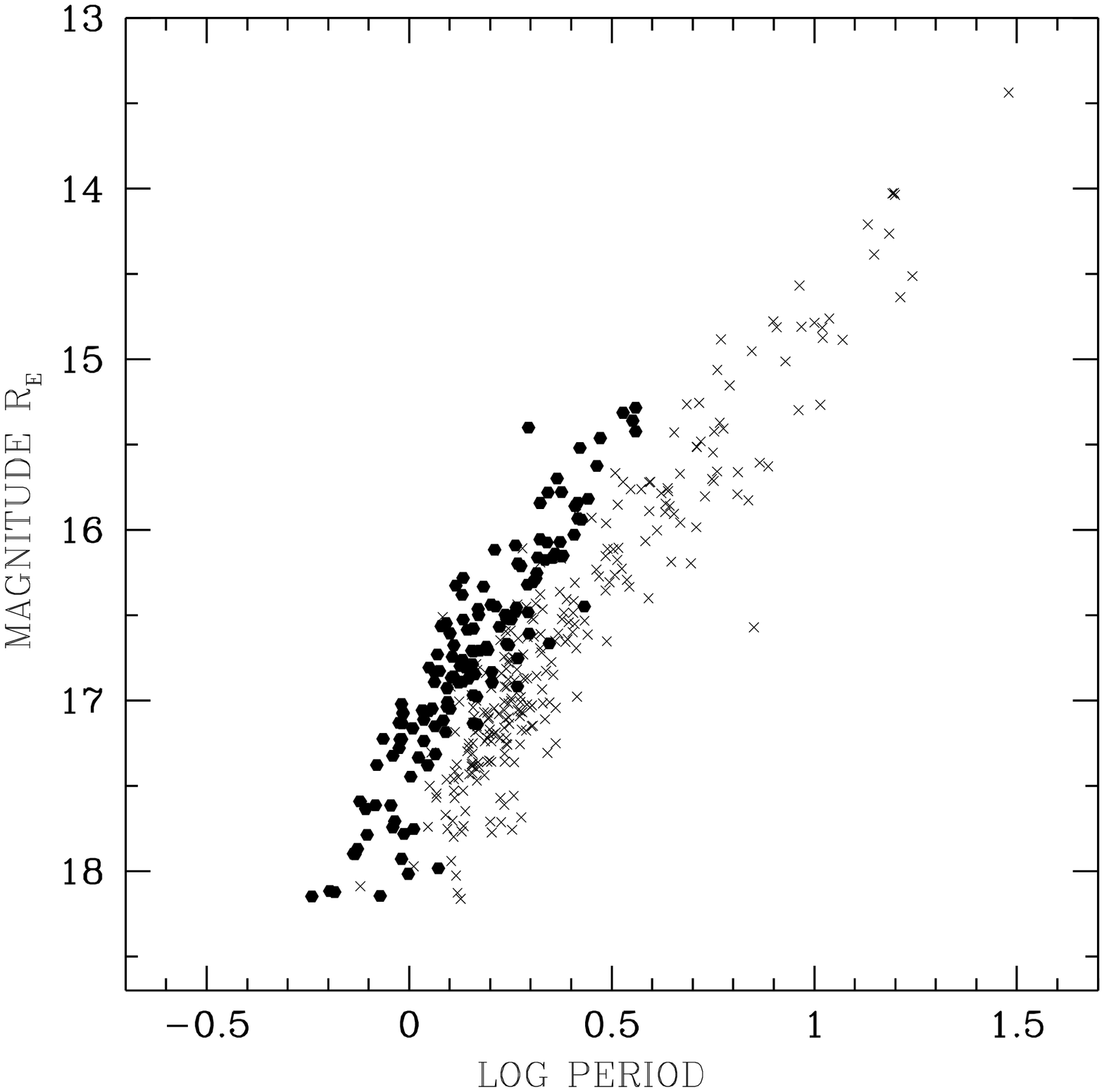,width=8.0cm} }  
  \caption{The PL relations for F and 1-OT Cepheids seen
in the EROS~1 database. The magnitudes are given in the EROS~1 filter system.
}\label{eros1}
\end{figure}
In addition, we looked directly in the SMC microlensing 
database of EROS~2 (\cite{palanque}) for the non--linearity.
This comparison was done for one CCD with a high density of Cepheids
(102 F Cepheids and 57 1-OT Cepheids).
The slope change is still visible. 
We also re-examined the PL relation for SMC F~Cepheids 
in the EROS~1 catalog (\cite{beaulieu96}; see Fig.~\ref{eros1}).
The slope change is visible. In the EROS~1 analysis
this phenomenon was ignored because of its marginal statistical significance.
Note that the EROS~1 data~set was obtained from measurements 
with another telescope, another CCD array, another photometry package 
and a different periodicity search program, so that observing the effect
in both the EROS~1 and EROS~2 data~sets seems a compelling argument
against most biases.

These two data~sets are the only published or available catalogs 
obtained from CCD observations that contain a significant 
number of short-period Cepheids ($P < 2$~days) towards the SMC.
The only previously published catalog with sensitivity to short period
SMC Cepheids (\cite{payne}) was obtained using photographic plates; 
no slope change is visible in it. However
as none of the possible observational biases considered above is able 
to explain away the effect observed in our data, we consider
the non--linearity established.
We observe (Fig. \ref{r21}) 
that this non--linearity occurs for periods similar to those
where the $R_{21}-\log P$ diagram exhibits a maximum ($R_{21}$ is the
amplitude ratio between the second and first harmonic).
\begin{table}[h]
\begin{tabular}{ll|c|c|c|c}
\hline
\multicolumn{2}{c|}{Populations} 
                  &  F       &    F        &   1-OT    &   1-OT       \\
     &            & P$<$2 d  &  P$>$2 d    & P$<$1.4 d & P$>$1.4 d   \\
\hline
 F   & P$<$2 d    &    --    &    0.0005   &    0.9689 &   0.0003    \\
 F   & P$>$2 d    &  0.0005  &     --      &    0.0049 &   0.5546    \\
1-OT & P$<$1.4 d  &  0.9689  &    0.0049   &      --   &   0.0034    \\
1-OT & P$>$1.4 d  &  0.0003  &    0.5546   &    0.0034 &    --       \\
\hline
\end{tabular}
   \caption{Kolmogorov-Smirnov probabilities from the comparison of 
SMC Cepheids distributions along the Wesselink $\eta$ coordinate. 
A small probability indicates different parent distributions. 
Similar tests along $\xi$ yield probabilities always larger than 15\%.
}\label{table_kolmogorov}
\end{table}
\section{Spatial distribution}\label{spatial}
To further investigate the effect, we have compared the
spatial distributions along the Wesselink $\eta$ and $\xi$ coordinates 
(\cite{wesselink}) of SMC F~Cepheids with 
$P < 2$~days and $P > 2$~days. 
\begin{figure}[htb]
\hbox{\epsfig{file=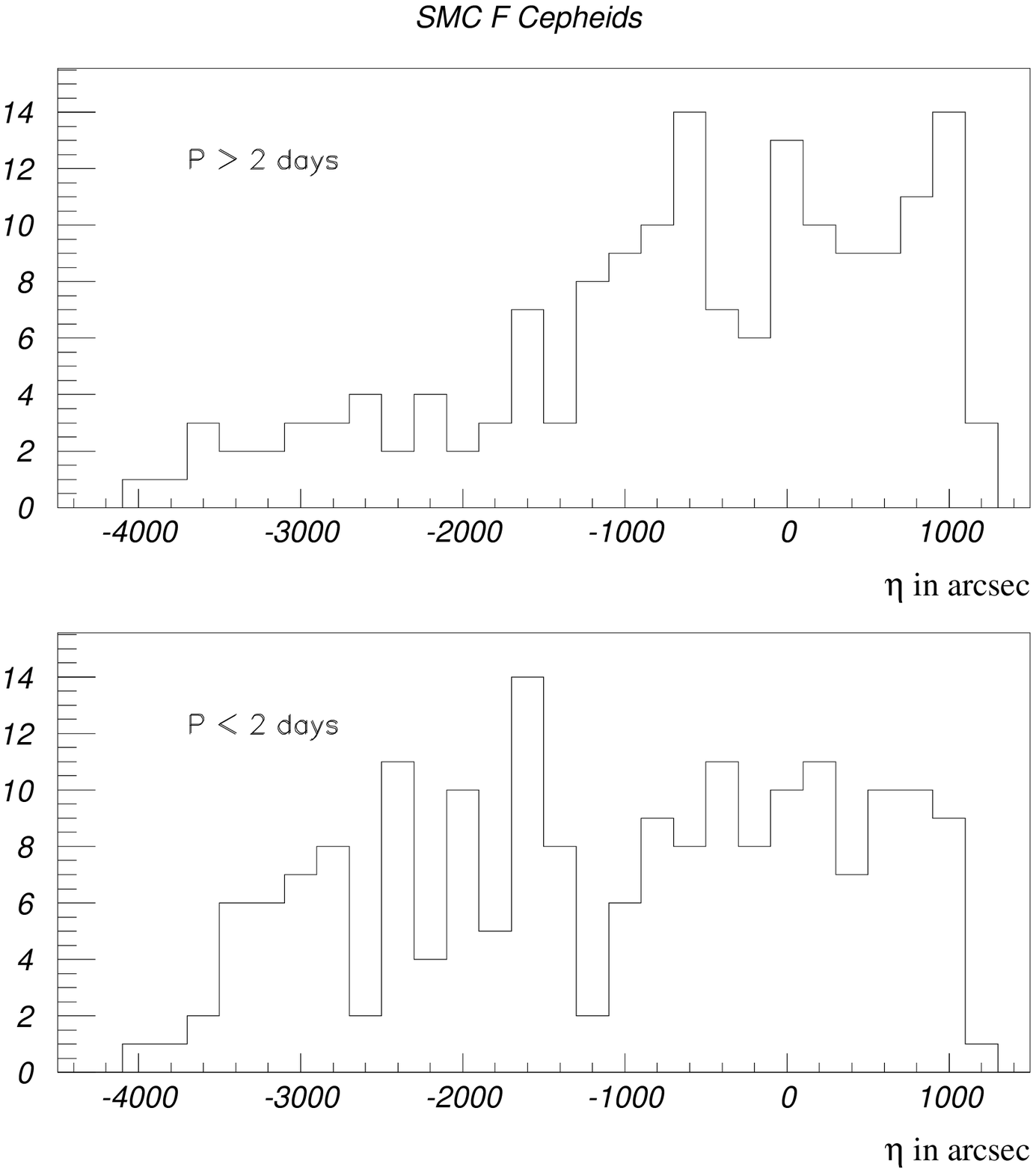,width=8.51cm} }  
\hbox{\epsfig{file=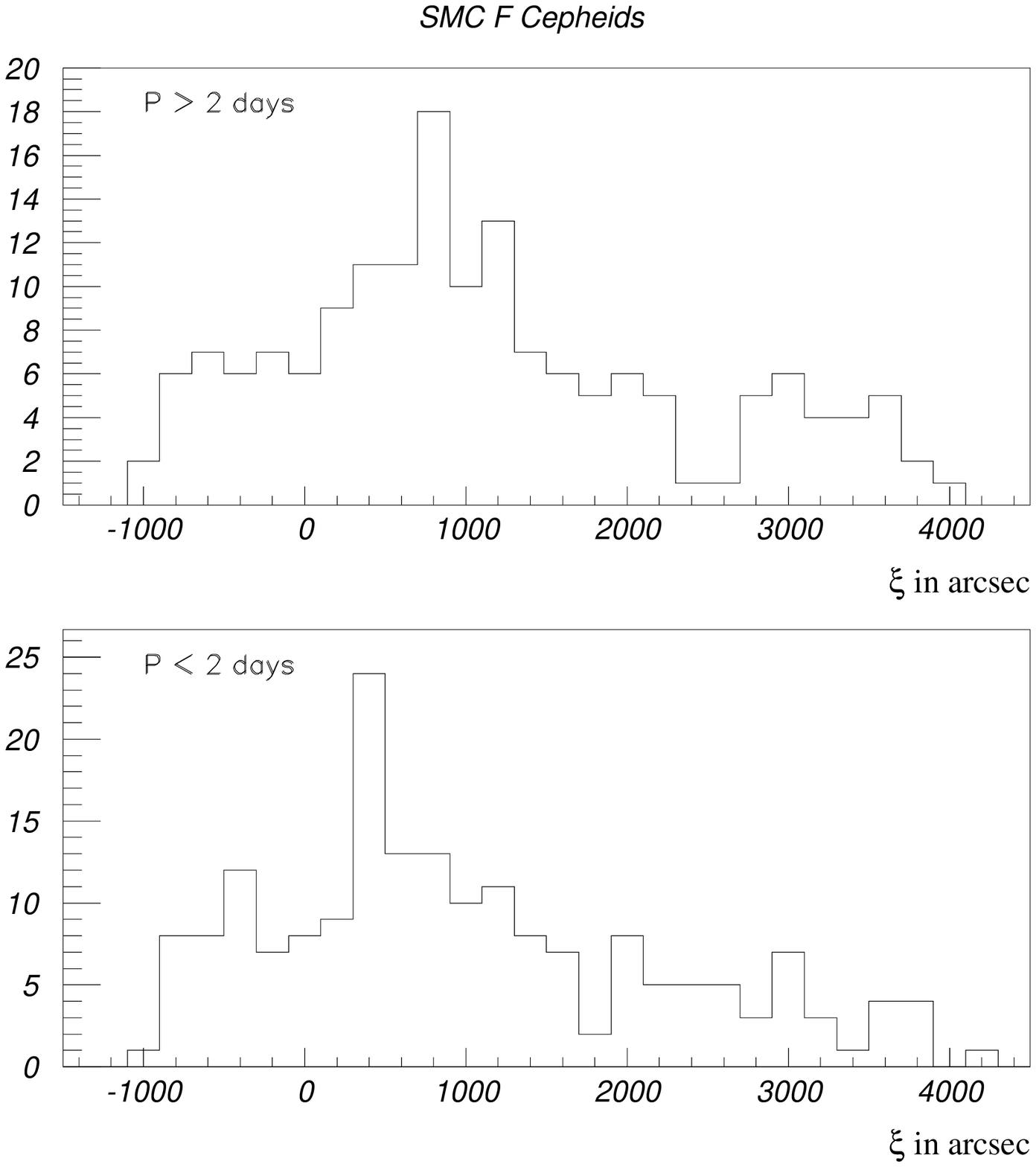,width=8.5cm} }  
\caption{The spatial distribution of SMC F Cepheids 
along the Wesselink $\eta$ and $\xi$ coordinates.}\label{distri_spatial}
\end{figure}
Along the $\xi$ coordinate, no significant difference
is observed. Along the $\eta$ coordinate, while SMC F~Cepheids with $P < 2$~days 
are uniformly distributed in our field, those with $P > 2$~days are concentrated
toward the center of the cloud (see Fig. \ref{distri_spatial}). 
A comparison between SMC 1-OT~Cepheids with $P < 1.4$~day and $P > 1.4$~day
leads to a similar conclusion. These observations are quantified
using a Kolmogorov-Smirnov test (see Table \ref{table_kolmogorov}).
From this table, it also appears that the distributions are
similar for short period F and 1-OT~Cepheids, and for long
period F and 1-OT Cepheids as well. To summarize, the distribution along $\eta$
of short period F and 1-OT Cepheids are incompatible with that of long
period Cepheids at the 4.5 $\sigma$ level at least.\\
In view of this, we have repeated the fits of Sect. 3 separately
for $\eta < -1500$ and $\eta > -1500$~arcsec, to search for a
possible spatial dependence of the slope change. This study is
inconclusive however because of the much lower abundance of long--period
Cepheids for $\eta < -1500$~arcsec (see top of Fig. \ref{distri_spatial}).
\begin{figure}
\hbox{\epsfig{file=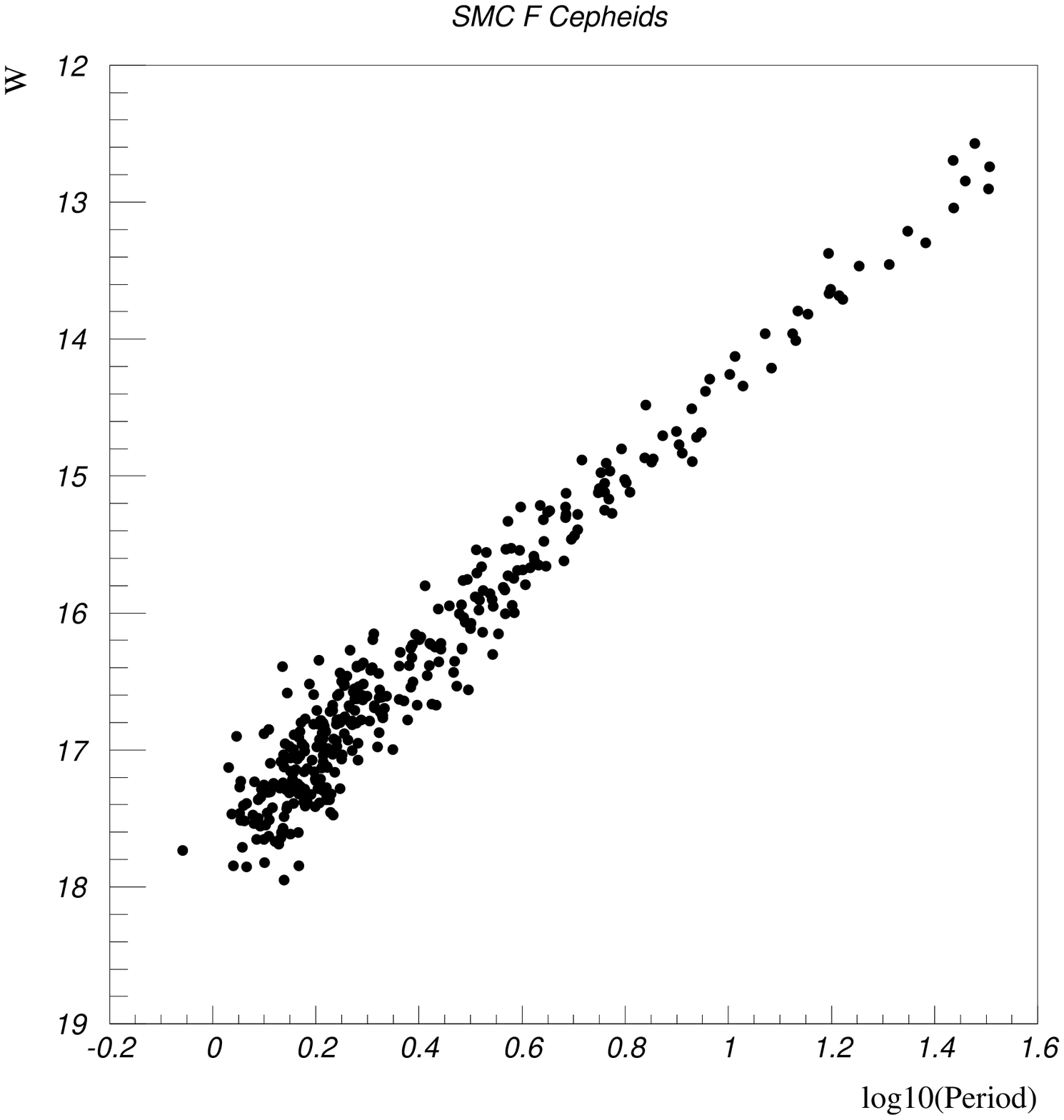,width=9cm} }  
\caption{The PL--relation as seen in W.}\label{wpl}
\end{figure}
\section{Probing the depth dispersion}
We have checked if the reported effect could be due to a distance effect,
using the reddening free magnitude $W$, corresponding to:
\begin{eqnarray}\label{wdef}
W = V_{\rm EROS} - A * (V_{\rm EROS} - R_{\rm EROS}) \\
A = R_{V_{\rm EROS}} / (R_{{V_{\rm EROS}}} - R_{R_{\rm EROS}}) = 3.44
\end{eqnarray}
The two factors $R_{{V_{\rm EROS}}}$ and $R_{R_{\rm EROS}}$ have been
calculated using a convolution of the EROS transmission function with
the extinction law given by \cite{cardelli}.\\
As observed by \cite{graff}, the reddening index $A$ in Eq. (\ref{wdef}) happens to 
be close to the color term $\gamma$ in the Period-Luminosity-Color equation:
\begin{equation}\label{plcdef}
V_{\rm EROS} = \alpha + \beta \log(P) + \gamma (V_{\rm EROS} - R_{\rm EROS})
\end{equation}
Subtracting Eq. (\ref{plcdef}) from Eq. (\ref{wdef}) yields:
\begin{equation}\label{wstraight}
W \sim \alpha + \beta \log(P) 
\end{equation}
\begin{figure*}
\hbox{\epsfig{file=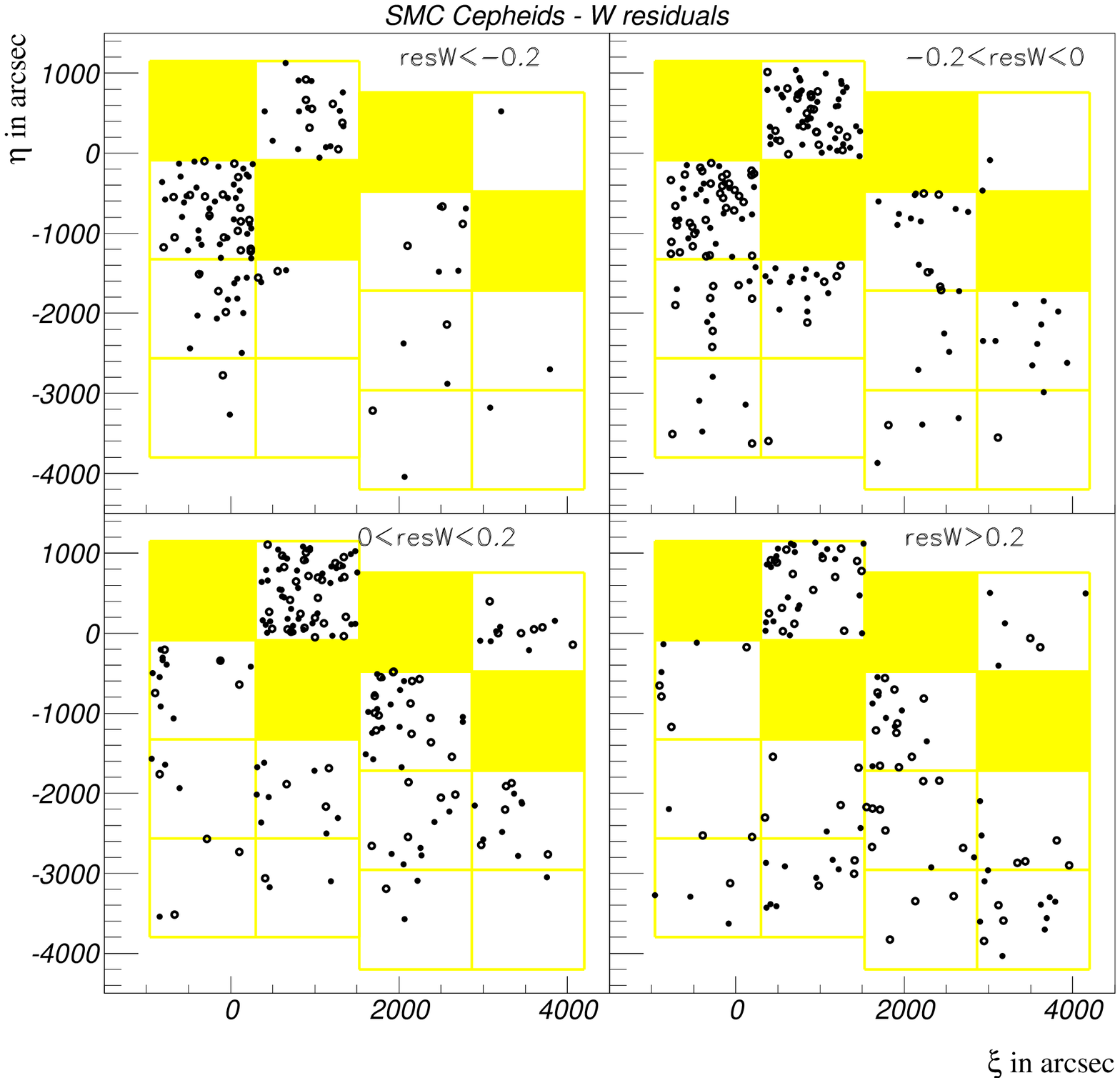,width=9cm} \epsfig{file=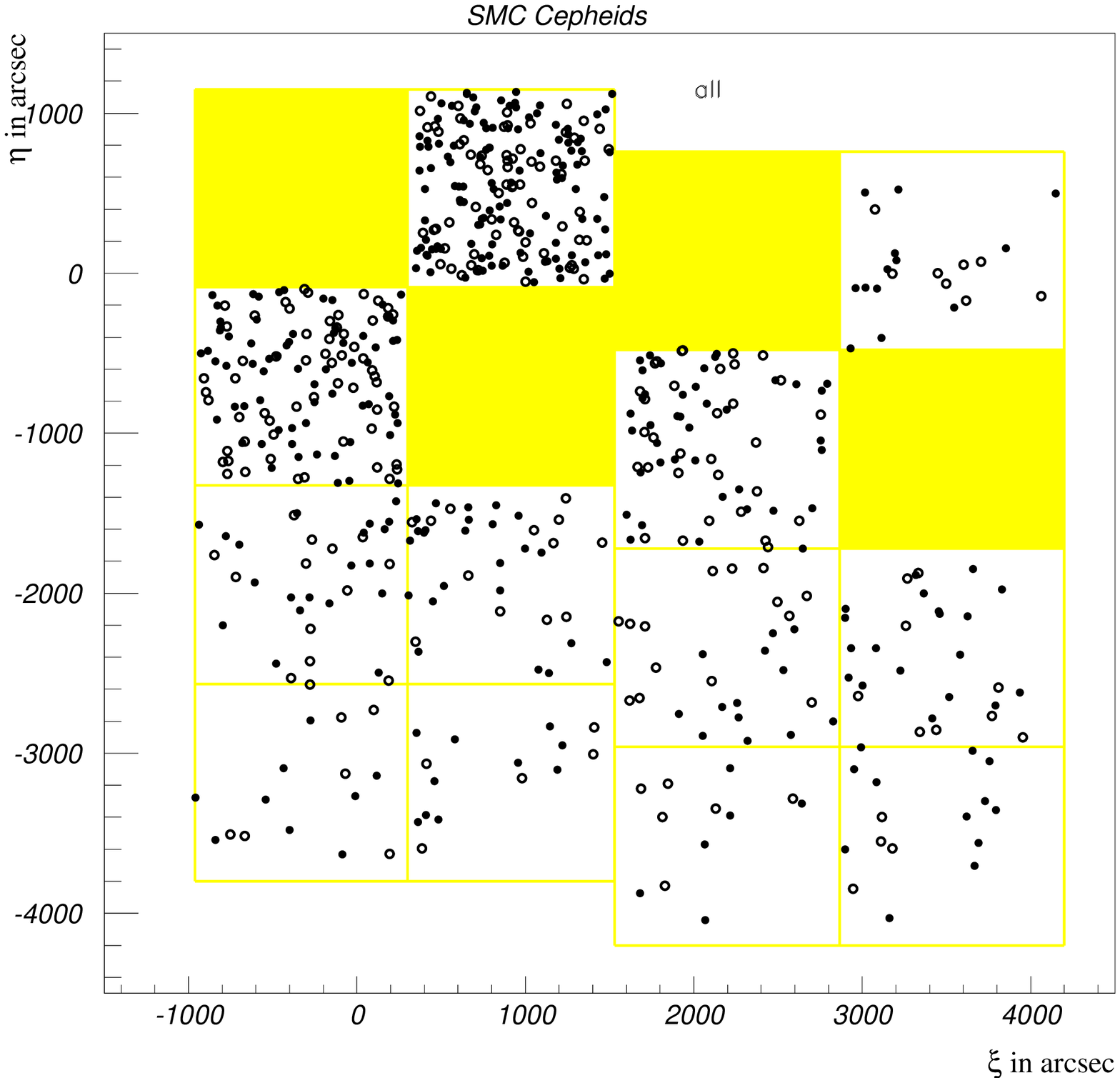,width=9cm}}  
\caption{The spatial distribution in Wesselink coordinates of the SMC F~Cepheids (black dots)
and 1-OT Cepheids (circles) for different bins of W--residuals.
The two types of Cepheids follow almost the same spatial distribution.
The two missing CCDs are indicated by shaded areas.}\label{resspatial}
\end{figure*}
Thus $W$ is almost free from the effects of reddening and of color in the instability strip.

As shown in Fig. \ref{resspatial} a difference in spatial
structure is visible for different intervals of $W$--residuals
\footnote{Residuals of W with respect to a $W-\log(P)$ linear fit.}, that
is probably a translation of the three dimensional structure of the SMC.

In Fig. \ref{wpl} we show the PL relation for F~Cepheids in the $W$ band.
The effect, though still visible, is not statistically significant.
Note however that the photometric error of 0.04~mag for $V_{\rm EROS}$
and $R_{\rm EROS}$ translate into a 0.15~mag dispersion in $W$.
Thus, given our photometric accuracy, we are unable to 
exclude a correlation between the observed non--linearity and
depth dispersion using this approach.
\section{Discussion}
It is beyond the scope of the present paper to give a firm explanation 
of the reported effect. We discuss below four different hypotheses based on 
qualitative considerations, which ideally should explain:
\begin{itemize}
\item the observed non-linearity of the PL relation for SMC F~Cepheids, 
\item the absence of non-linearity for SMC 1-OT Cepheids, 
\item the spatial distributions mentionned in Sect.~4.  
\end{itemize}
They are:
\begin{itemize}
\item a superposition of two Cepheid populations of different ages at different distances;
\item the mixing of two classes of variable stars;
\item a non uniform filling of the instability strip;
\item a thinning of the instability strip itself at short periods.
\end{itemize}
\subsection{Superposition of two Cepheid populations}
In the previous section we have seen that long and short period
Cepheids are distributed differently along the $\eta$ direction.
If their distributions differ also along the line-of-sight, and if
the short period Cepheids are in average more distant, the
slope variation in the PL relation could be explained.
For example a population of short period (older) Cepheids behind
a population of long period (younger) Cepheids could indicate
a star formation region in the foreground.
In this case the population of short period Cepheids would be remnants of 
older star formation regions and thus exhibit a flatter spatial distribution
along the $\eta$ coordinate.  
On the other hand the absence of a non-linearity for 1-OT Cepheids, in spite of the fact that
they have similar $\xi$ and $\eta$ distributions as F~Cepheids, challenges this explanation. 
To further investigate this, we will soon extend the present
study to a wider field. Confirmation of this scenario would provide
a tool to study the recent history of star formation in the SMC.
\subsection{Mixing of two classes of variable stars}
This hypothesis involves the superposition of two stellar populations,
classical Cepheids and the so-called anomalous Cepheids
\footnote{Note that we have excluded W Virginis stars in the selection process.}.
The latter have periods ranging between $\approx$ 0.5---2 days and 
are seen in dwarf spheroidal galaxies.
Their PL relation is known to be metallicity dependent (\cite{nemec_etal}).
For a typical SMC metallicity, anomalous Cepheids would be
too faint for a superposition to occur. Using the
PL relations of classical Cepheids (Madore \& Freedman 1991) and
anomalous Cepheids (\cite{nemec_etal}), we estimate
that a superposition requires values of [Fe/H] as low as -3.3 $\pm$ 0.6 for 
a 1 day period in the B passband, well below common SMC metallicities.
%{\bf JPB a enlever: 
%Spectroscopic studies would be needed to confirm if the observed slope change 
%is due to a new class of extremely metal-poor variable stars.
%Furthermore these metal-poor variable stars should have had the time to
%mix up and thus give the flat spatial distribution seen along the $\eta$ coordinate.}
%
\subsection{Stellar evolution; a non uniform filling of the instability strip}
As detailed in \cite{madore}, 
the PL relation is the projection of a narrow band in
the period--luminosity--colour plane called the
instability strip, where stellar envelopes become unstable and are subject 
to radial pulsations.
During their evolution, stars in the mass range 2--8 M$_{\sun}$ 
will typically cross this zone three times. The first crossing occurs when stars 
evolve rapidly away from the main sequence. The two subsequent crossings 
occur when 
these stars start helium burning and evolve on horizontal evolution tracks 
(or blue loops). The extent of the blue loop is very sensitive to metallicity and 
will tell whether a star of given mass will cross the instability strip or not. 
It is also known that this loop becomes smaller with decreasing mass.

In this scenario
(Fig. \ref{scenario}, scenario 3)
the blue loop for stars with masses 
smaller than $\sim$3.6 M$_{\sun}$ (corresponding to $P \approx 2$ days SMC Cepheids) 
would stop before reaching the blue edge of the instability strip.
The resulting observed non--linearity of the PL relation, below a given period,
would be the consequence of this decreasing extent of blue loops
at small masses.
Following the observations reported here and in \cite{bauer1}, 
Baraffe et al. (1998) have computed evolutionary models, 
coupled to a linear non-adiabatic stability analysis, in order
to get consistent mass--period--luminosity relations.
They report a change in the slope consistent with our observations.
 
On the other hand this hypothesis has to fulfill one important constraint : the instability strip 
for 1--OT Cepheids is in average bluer than that for F~Cepheids; thus a 
depopulation of the 1--OT instability strip should also be visible, 
at about 0.7 $\times$ 2~days for SMC Cepheids. The absence of such a 
clear slope change for SMC 1--OT Cepheids challenges this explanation.
%{\bf JPB a enlever:
%Futhermore this hypothesis does not explain the different spatial distributions
%for long and short period SMC Cepheids.}
%
\subsection{ Pulsation theory; the shape of the instability strip}
\begin{figure}
\hbox{\epsfig{file=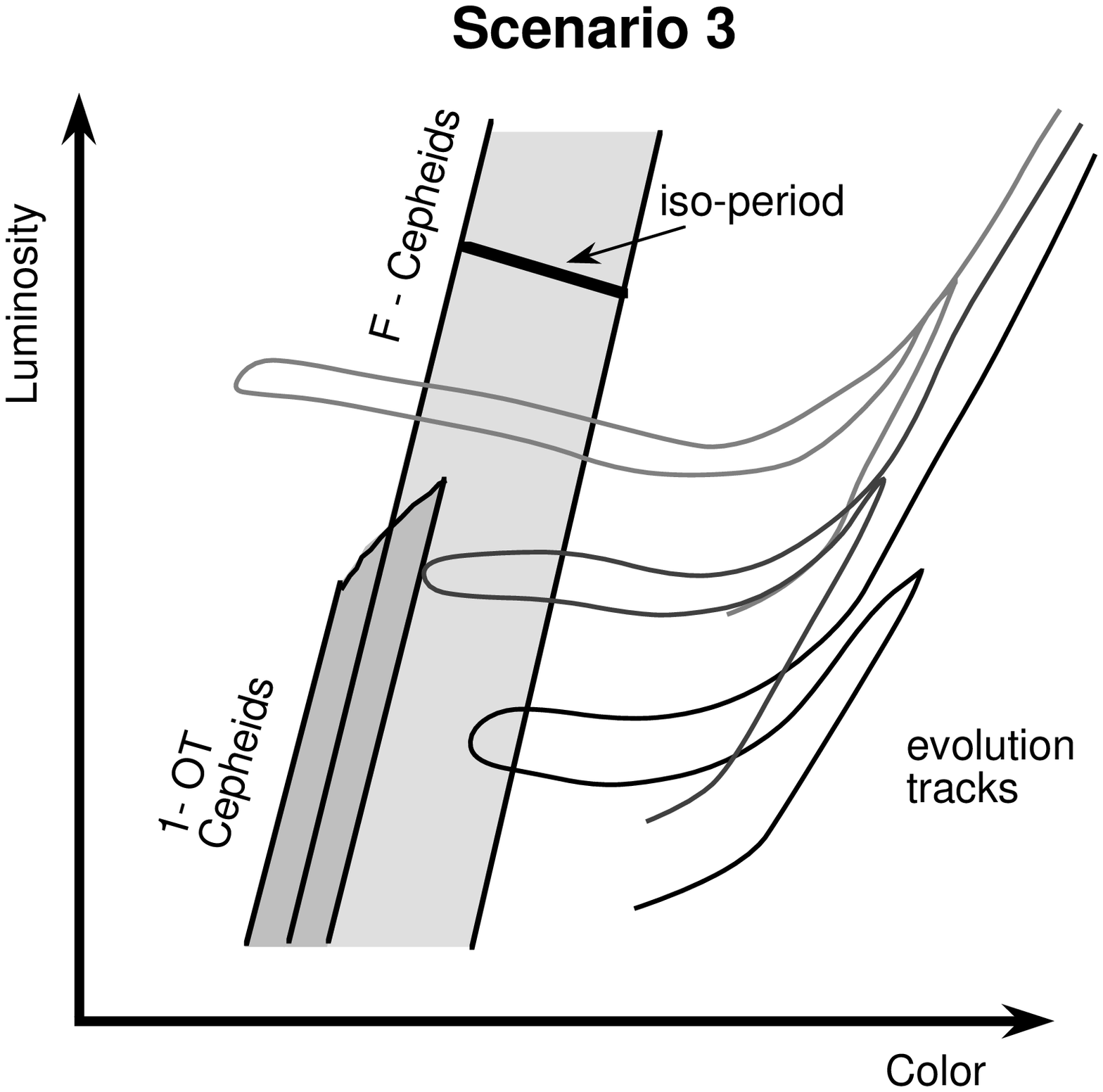,width=7.0cm} }  
\hbox{\epsfig{file=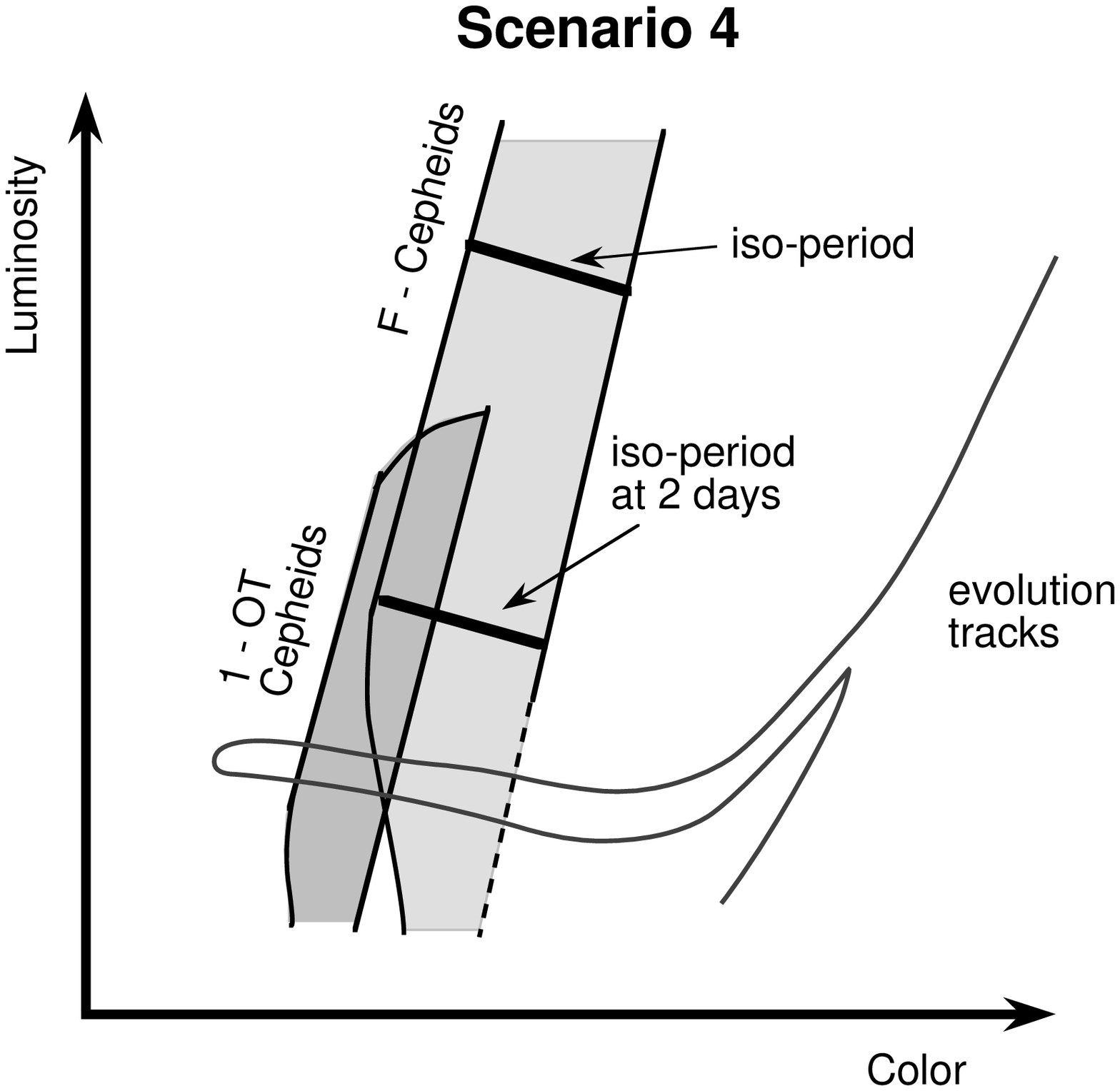,width=7.0cm} }  
\caption{Two possible explanations of the reported phenomena.
In the third scenario we are witnessing a gradual depopulation 
of the instability strip
due to the fact that horizontal evolutionary tracks 
become smaller with decreasing mass or period. 
In the fourth scenario the blue edge of the instability strip could be
non--linear.}\label{scenario}
\end{figure}

This hypothesis (Fig. \ref{scenario}, scenario 4) 
involves the physics of the pulsation itself 
and assumes that the blue edge of the instability strip 
becomes non--linear around $P = 2$~days for F~Cepheids. 
In this admittedly speculative scenario, the absence of an effect for
1~OT Cepheids seems possible.
The observed non-linearity could thus give  
a new and useful constraint for stellar pulsation 
theory describing short period (i.e. small mass) and 
low metallicity Cepheids. 
%{\bf JPB A enlever: As for scenario 3, the distinct spatial
%distribution is not explained.}
%
\subsection{LMC Cepheids}
It is worth closing this discussion with remarks about LMC Cepheids
in the light of the four scenarios discussed above.

Scenario~1 is specific to the SMC and has no implication for LMC Cepheids.

Regarding scenarios 2 and 3,
in the interesting mass range (2---4 M$_{\sun}$), 
stellar evolution theory predicts smaller blue loops for the 
LMC than for the SMC because of the higher LMC metallicity. Thus 
a non uniform population of the instability strip 
(scenario 3) should also be visible in our
LMC~F Cepheid dataset, but for even longer period Cepheids.
An indication for an abrupt end of the
PL relation for LMC~F Cepheids can be seen in Fig. \ref{pl}
at 2.5 days; this has also been observed and modelled by the
{\sc macho} collaboration (\cite{alcock}), who explain the short period
Cepheids as beeing anomalous Cepheids.
In our sample the 12 LMC F~Cepheids with $P < 2.5$~days 
and 37 LMC 1-OT Cepheids with $P < 1.7$~days,
would belong to this anomalous Cepheid population.

In scenario 4 finally, we would not observe a non--linearity of the
blue edge of the instability strip, if stellar evolution constrains
the population of F~Cepheids to periods larger than 2.5~days.
\section{Conclusion }
We have presented the PL diagrams from a dedicated
LMC/SMC Cepheid observation campaign with the EROS~2 setup. 
For the first time, a change in the slope of the PL relation for
short period SMC F~Cepheids is observed. For SMC F~Cepheids 
with periods smaller than 2.0 days the magnitude deviation compared 
to the entire PL relation reaches 0.2~mag.
On the other hand we observe no similar effect for 1-OT Cepheids.
Furthermore, long and short period SMC Cepheids display significantly different
spatial distributions. We have presented different possible
explanations for these observations and it is clear that
further studies are needed to select the most likely one.
With the amount of EROS data available and if the absence of
a slope change for 1-OT Cepheid is confirmed, scenario 4 
--~a curvature of the blue edge of the instability strip~-- seems the most natural one.
Finally, we remark that the reported phenomenon does not affect the 
PL relation used in the HST Key project for the determination 
of H$_{ \rm 0}$, as it is based on longer period Cepheids.
\begin{acknowledgements}
We thank the referee, whose remarks led us to 
look closer into the spatial distributions.
The authors are particularly grateful to the ESO staff 
at the La Silla Observatory for their night and day assistance.
We wish to thank the Observatoire de Haute Provence technical staff 
for their help in refurbishing and mounting 
the Marly telescope. We are grateful to the DAPNIA technical
staff for the maintenance of the 2 CCD cameras.
We wish to thank J.F. Lecointe for assistance with the online
computing and the CCIN2P3 staff, for their help 
during this first EROS~2 massive data production. 
\end{acknowledgements}
{\bf Note added:} \\
After this paper was submitted, the {\sc OGLE } collaboration
presented results on beat and second--overtone Cepheids in the SMC based
on a larger sample (\cite{udalski1}, \cite{udalski2}). 
We observe good aggreement betweeen their
$R_{21}-\log P$ diagram and ours, and remark that their data sample could
be used to confirm or study further the non-linearity that we have reported here.
\end{document}